\documentclass[journal=jacsat,manuscript=article]{achemso}
\usepackage{graphicx}
\usepackage{color}
\usepackage{titlecaps}
\usepackage{caption}
\usepackage{amssymb}
\usepackage{amsmath}
\usepackage{gensymb}
\usepackage{booktabs}
\usepackage{multicol}
\usepackage[table]{xcolor}
\usepackage{etoolbox}
\makeatletter
\def\acs@contact@details{
E-mail: \acs@email@list
}
\makeatother

\title{Predictive quantum vibrational spectra through active learning 4G-NNPs}

\setkeys{acs}{
  abbreviations = false,
  articletitle  = true,
  biblabel      = brackets,
  biochem       = false,
  doi           = false,
  etalmode      = truncate,
  keywords      = false,
  maxauthors    = 95,
  super         = true
}

\author{Md Omar Faruque}
\author{Dil K. Limbu}
\author{Nathan London}
\author{Mohammad R. Momeni}
\email{mmomenitaheri@umkc.edu}
\affiliation{Division of Energy, Matter and Systems, School of Science and Engineering, University of Missouri $-$ Kansas City, Kansas City 64110, MO, United States}

\abbreviations{}

\begin{document}

\begin{abstract}
Predictive simulation of vibrational spectra of complex condensed-phase and interface systems with thousands of degrees of freedom has long been a challenging task of modern condensed matter theory. In this work, fourth-generation high-dimensional committee neural network potentials (4G-HDCNNPs) are developed using active learning and query-by-committee, and introduced to the essential nuclear quantum effects (NQEs) as well as conformational entropy and anharmonicities from path integral (PI) molecular dynamics simulations. Using representative bulk water and air-water interface test cases, we demonstrate the accuracy of the developed framework in infrared spectral simulations. Specifically, by seamlessly integrating non-local charge transfer effects from 4G-HDCNNPs with the NQEs from PI methods, our introduced methodology yields accurate infrared spectra using predicted charges from the 4G-HDCNNP architecture without explicit training of dipole moments. The framework introduced in this work is simple and general, offering a practical paradigm for predictive spectral simulations of complex condensed phases and interfaces, free from empirical parameterizations and ad hoc fitting.
\end{abstract}

%%%MAIN TEXT%%%%
\section{Introduction}  \label{sec1}
%\textbf{Introduction\newline} 
Vibrational spectroscopy is an excellent tool for probing complex processes across different condensed-phase and interface environments.\cite{cr_116_7590,nl_8_2959,pccp_13_17658,pccp_16_12202} To obtain accurate predictive vibrational spectra, free from \textit{ad hoc} fittings or empirical parameterizations, it has been well-established that robust dynamical methodologies that incorporate anharmonicities and conformational entropies, as well as the essential nuclear quantum effects (NQEs) are needed to be developed and integrated with accurate potential energy surfaces (PESs) that include non-local charge transfer interactions.\cite{ceriotti2016nuclear, markland2018nuclear} Path integral molecular dynamics (PIMD) offers an accurate and efficient route to include NQEs by representing a quantum particle as a series of identical classical particles, referred to as beads connected via harmonic springs to form a closed loop, referred to as a ring polymer.\cite{julFeynman2010,janParrinello1984,Hirshberg:2019} Accurate equilibrium properties that incorporate NQEs can be calculated using PI Monte Carlo (PIMC) or PI molecular dynamics (PIMD) simulations.\cite{janParrinello1984} Similarly, approximate real-time PI methods, including thermostatted ring polymer MD (TRPMD),\cite{junRossi2014} partially adiabatic centroid MD (PA-CMD)\cite{jcp_124_154103}, and a myriad of methods branched from them,\cite{janHone2005, qcmd, musil2022quantum, limbu2025hcmd}, can then be used for the simulation of dynamical properties including vibrational spectra through the calculations of appropriate time correlation functions with NQEs included. 

As mentioned above, accurate PESs that incorporate long-range charge transfer interactions in condensed phases and interfaces are crucial for predictive spectral simulations. Traditional empirical force fields are computationally efficient, but they lack the accuracy and transferability required to capture the complex processes involved in condensed phases and interfaces. On the other hand, \textit{ab initio} PIMD simulations where forces felt by nuclei are calculated on-the-fly using electronic structure methods are prohibitively expensive for obtaining converged real-time dynamical properties, especially at low temperatures.\cite{marsalek2017quantum} Machine learning interatomic potentials (MLIPs) have recently been shown to be able to bridge the gap between the accuracy and the cost of these simulations.\cite{behler2016perspective} This is due to the ability of MLIPs in leveraging data-driven approaches to approximate PESs with near \textit{ab initio} accuracy at a fraction of the computational expense.\cite{thiemann2024introduction} One of the early established MLIP architectures that uses artificial neural networks (ANNs) is the high-dimensional neural network potentials (HDNNPs), which are widely adopted to construct PESs for both gas-phase and condensed-phase molecular systems.\cite{behler2007generalized}
\begin{figure*}[!t]
    \centering
    \includegraphics[width=0.99\linewidth]{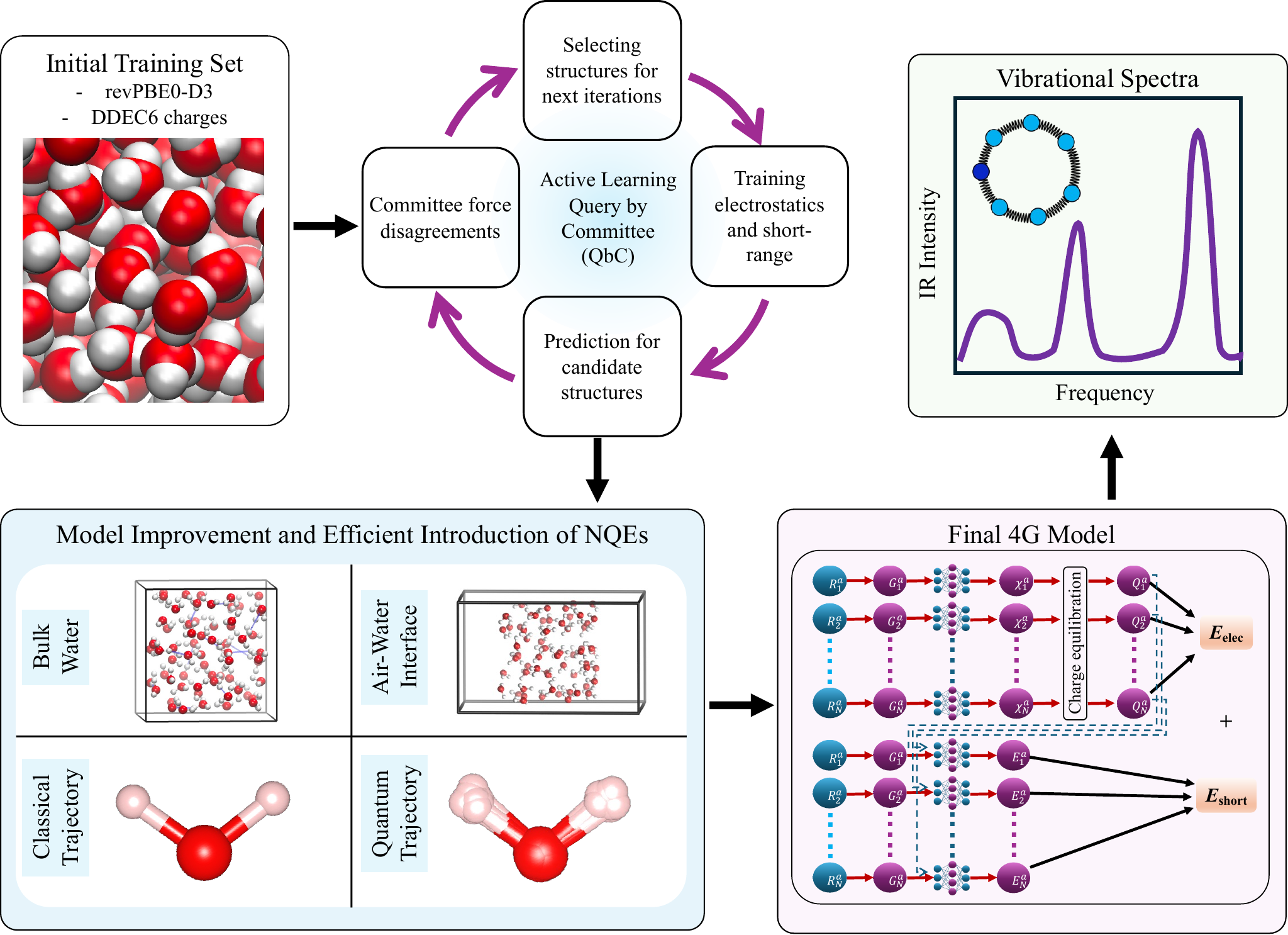}
    \caption{The adopted workflow for 4G-HDCNNPs training and vibrational spectral simulations using modified AML\cite{schran2020committee} with NQEs incorporated.}
    \label{fig:workflow}
\end{figure*}
Early generations of HDNNPs primarily focus on local atomic environments and are shown to excel in capturing short-range interactions. However, they fail to accurately represent non-local interactions, which are crucial for systems exhibiting significant charge transfer or polarization effects.\cite{shaidu2024incorporating} 

%6. 
To address this limitation, fourth-generation high-dimensional neural network potentials (4G-HDNNPs) have recently been introduced.\cite{ko2021fourthgeneration} 4G-HDNNPs integrate environment-dependent atomic energies with long-range electrostatics from learned atomic charges, enabling a robust description of complex environments.\cite{ko2021fourthgeneration} More specifically, the 4G architecture employs a charge equilibration scheme to determine atomic charges that reflect the global electronic structure. This approach allows for the inclusion of non-local phenomena such as long-range charge transfer relevant in redox chemistry.\cite{gubler2024accelerating, kocer2024machine} A key advantage of 4G-HDNNPs is their inherent ability to predict and provide atomic charges, once appropriately trained on \textit{ab initio} reference charges, a feature that is lacking in previous generations. Therefore, the charges predicted by 4G-HDNNP-enabled dynamics simulations can, in principle, be directly utilized for simulating infrared (IR) spectra through the calculation of dipole autocorrelation functions, without the explicit training of dipoles. Alternative hybrid methods, such as CombineNet\cite{tu2025longrange}, also incorporate machine-learned short-range energies with explicit long-range electrostatics. Such methods fundamentally differ from 4G-HDNNP as their short-range component does not include any charge information, whereas 4G-HDNNPs add charge as an input layer for the short-range neural networks as well.\cite{ko2021fourthgeneration}

%7. 
In this work, 4G-HDNNPs are developed and integrated with different path integral methods for the accurate prediction of the IR spectra of representative bulk water and air-water interface systems. Water’s complex structure arises from its dynamic hydrogen-bond (H-bond) network, with differential intra- versus inter-molecular H-bonds, which significantly influence its vibrational spectral properties.\cite{elton2016hydrogenbond} Recent studies have shown that this complex network supports propagating optical phonon-like modes, indicating coherent long-range interactions.\cite{elgabarty2020energy} These long-range effects play a crucial role in water’s vibrational dynamics, affecting energy transfer processes and the overall behavior of its H-bond network. To achieve fast and efficient spectral calculations, an active learning framework using query-by-committee (QbC) is developed and employed to facilitate training of the 4G-HDNNP models. Our QbC approach enables data-efficient training through an iterative selection of the most distinct and informative configurations. Thus, we refer to our developed potentials 4G-HDCNNPs, where "C" stands for committee.

%\textbf{Previous attempts to combine HDNNPs+PIMD\newline}
Although attempts have been made to combine HDNNPs with PIMD simulations before,\cite{shiga2025computation} this work is the first report to directly employ fourth-generation PI simulations for the calculations of IR spectra, bypassing the need for the explicit training of the dipoles. The 4G-HDCNNPs reported here have two pivotal advances compared to the second-generation HDNNP (2G-HDNNP) models. First, 4G-HDCNNPs incorporate explicit non-local charge-transfer interactions, capturing interactions that 2G models neglect and thereby delivering higher accuracy in energies and forces for environments with significant charge transfer and polarization effects.\cite{ko2021fourthgeneration, kocer2024machine, kocer2025iterative} Second, as the 4G architecture predicts atomic charges, charge-dependent observables such as dipole moment can be evaluated directly, without the approximations that 2G models require. Below, we first provide a brief theoretical background for 4G-HDCNNPs, along with details of model training and IR spectral simulations, followed by our simulated IR spectra for the considered bulk water and air-water interface systems.

%%%%%%%%%%%%%%%%%%%%%%%%%%%%%%%%%%%%%%%%%%%%%
% SECTION 2
%%%%%%%%%%%%%%%%%%%%%%%%%%%%%%%%%%%%%%%%%%%%%
\section{Theory Details} \label{sec2}
%\noindent\textit{Theory Details\newline}
%\textit{4G-HDNNPs\newline}
\subsection{4G-HDNNPs}
4G-HDNNPs seamlessly combine atom-centered descriptors with environment-dependent charges obtained from a global charge equilibration, taking into account both local short-range interactions and non-local long-range electrostatics. This is achieved by employing one set of neural networks to predict environment-dependent electronegativities for long-range electrostatics and a second set of neural networks to model the local, short-range contributions. The schematic of the networks and their interrelation is depicted in Fig. \ref{fig:workflow}. This enables the potential to respond to non-local charge transfer and polarization without sacrificing accuracy in capturing local interactions. The total energy in a 4G-HDNNP is, therefore, expressed as\cite{ko2021fourthgeneration}
\begin{equation}
    E_{\text{total}}(R, Q) = E_{\text{elec}}(R, Q) + E_{\text{short}}(R, Q),
    \label{eq-1}
\end{equation}
where $R$ and $Q$ denote atomic positions and total charge, respectively. The $E_{\text{elec}}$ term represents long-range electrostatic interactions based on atomic charges, while $E_{\text{short}}$ denotes short-range interactions. The electrostatic component $ E_{\text{elec}} $ is computed by combining pairwise screened Coulomb interactions and self-terms \cite{ko2021fourthgeneration}
\begin{equation}
    E_{\text{elec}} = \sum_{i=1}^{N} \sum_{j \neq i}^{N} \frac{\text{erf} \left( \frac{r_{ij}}{\sqrt{2} \gamma_{ij}} \right)}{r_{ij}} Q_i Q_j + \sum_{i=1}^{N} \frac{Q_i^2}{2\sigma_i \sqrt{\pi}},
    \label{eq-3}
\end{equation}
where $r_{ij}$ is the distance between atoms $i$ and $j$, and $\gamma_{ij} = \sqrt{\sigma_i^2 + \sigma_j^2}$ with $\sigma_i$ and $\sigma_j$ representing Gaussian widths. The error function, $\text{erf} \left( \frac{r_{ij}}{\sqrt{2} \gamma_{ij}} \right)$, smoothens the interaction between two atomic charges at a distance of $r_{ij}$.

Earlier third-generation HDNNPs predict charges directly from local descriptors, which limits their ability to equilibrate charge over extended networks.\cite{behler2021four} Therefore, to include the long-range charge transfer effects, a separate charge equilibration is performed in 4G-HDNNPs. In this scheme, the atomic charges are optimized by minimizing $E_{Q}^{\text{eq}}$\cite{ko2021fourthgeneration}
\begin{equation}
    E_{Q}^{\text{eq}} = E_{\text{elec}} + \sum_{i=1}^{N} \left( \chi_i Q_i + \frac{1}{2} J_i Q_i^2 \right),
    \label{eq-2}
\end{equation}
where $\chi_i $ are the learned atomic electronegativities and $J_i$ are the element-specific hardness. The optimization is performed by solving the following equation:
\begin{equation}
    \frac{\partial E_{Q}^{\text{eq}}}{\partial Q_i} = 0, \quad i = 1, \dots, N,
    \label{eq-4}
\end{equation}
which yields the set of linear equations
\begin{equation}
    \sum_{j=1}^{N} A_{ij} Q_j + \chi_i = 0.
     \label{eq-5}
\end{equation}
The $ A_{ij} $ matrix elements are defined for periodic systems using Ewald summation as \cite{kocer2025iterative}
\begin{equation}
    \begin{aligned}
        A_{ij} &= \frac{4\pi}{V}\sum_{k \ne 0}\frac{e^{-\frac{\eta^2|\mathbf{k}|^2}{2}}}{|\mathbf{k}|^2}\cos(\mathbf{k}\cdot\mathbf{r}_{ij}) \\ 
        &\quad+ \begin{cases}
            J_i - \frac{2}{\sqrt{2\pi}\eta} + \frac{1}{\sqrt{\pi}\sigma_i}, & i = j \\
            \\
            \frac{\mathrm{erfc}\left(\frac{r_{ij}}{\sqrt{2}\eta}\right) - \mathrm{erfc}\left(\frac{r_{ij}}{\sqrt{2}\gamma_{ij}}\right)}{r_{ij}}, &  i \ne j,\, r_{ij} < r_\text{cut}^{\text{real}},
        \end{cases}
    \end{aligned}
\end{equation}
where $\mathbf{k}$ is the reciprocal lattice vector, $V$ is the volume of the periodic simulation cell, and $\eta$ is a hyperparameter that defines the width of the Gaussian charges in the real-space cutoff within the Ewald summation. Summation over $\mathbf{k}$ is performed excluding the zero vector. The first term, or the reciprocal-space sum, represents long-range Coulomb interactions computed in the reciprocal space, and the second, or real-space interaction term, is accounted for by the complementary error function terms (erfc) to ensure proper convergence and efficiency. The real-space contribution is evaluated only for pairs with $r_{ij} < r_\text{cut}^\text{real}$, where $r_\text{cut}^\text{real}$ is the real-space cutoff.

Similar to the second and third-generation HDNNPs, the short-range energy is formulated as the sum of atomic energies
\begin{equation}
    E_{\text{short}} = \sum_{i=1}^{N} E_i,
    \label{eq-7}
\end{equation}
with $E_i$ representing the energy contribution of atom $ i $ predicted by the second neural network that takes the local symmetry functions and the atomic charge as input. To avoid double-counting, the electrostatic energy is subtracted from the reference-calculated total energy before training the second neural network.

\subsection{IR Spectral Simulations}
%\noindent\textit{IR Spectral Simulations}
Using 4G predicted charges directly and without explicit training of dipoles, IR spectra are calculated from the Fourier transform of the total cell dipole-derivative autocorrelation function, following our previous works,\cite{limbu2025hcmd, london2025dl_poly}
\begin{equation}
       \tilde{I}(\omega) = \frac{1}{2\pi}\int_{-\infty}^{\infty}dt\ \mathrm{e}^{-i\omega t} C_{
       \dot{\mathbf{M}}{\dot{\mathbf{M}}}}(t) f(t)
\end{equation}
where $\dot{\boldsymbol{M}}$ is the time
derivative of the total dipole moment, and $f(t)$ is a window function that dampens the tail of the
autocorrelation function and eliminates ringing artifacts in the calculated spectra.\cite{benson2020which} . Here we use the Hann window~\cite{press_etal:1992}
\begin{equation}
  f(t) = 
  \begin{cases}
    \cos^2\left( \frac{\pi t}{2\tau}\right)\quad &|t| \leq \tau \\
      0 & |t| > \tau
  \end{cases}  
\end{equation}
 where $\tau$ is a cutoff time chosen as appropriate.

%%%%%%%%%%%%%%%%%%%%%%%%%%%%%%%%%%%%%%%%%%%%%
% SECTION 3
%%%%%%%%%%%%%%%%%%%%%%%%%%%%%%%%%%%%%%%%%%%%%
\section{Simulation Details} \label{sec3}
%\noindent\textit{Simulation Details\newline}
\subsection{4G-HDCNNP Model Training}
%\textit{4G-HDCNNP Model Training}
The bulk water simulation box contains 64 water molecules in a cubic box of size 12.42~\AA, corresponding to a density of 0.997 g/cm$^{3}$ at 298 K. For the interfacial simulations, 96 water molecules were placed in a 14.21~\AA~$\times$ 14.21~\AA~$\times$ 39.21~\AA~ rectangular prism with a 12.5~\AA~ vacuum layer on each side of the slab along the $z$-direction. Periodic boundary conditions were used in all three directions. The reference \textit{ab initio} MD (AIMD) trajectories were generated employing the revPBE0-D3 dispersion corrected hybrid functional\cite{adamo1999reliable,grimme2010consistent,perdew1996generalized,zhang1998comment} in CP2K \cite{kuhne2020cp2k} using the Quickstep module\cite{vandevondele2005quickstep}. Reference atomic charges were calculated using the DDEC6 charge partitions.\cite{manz2016introducing} Previous studies have shown that NQE-included hybrid revPBE0-D3 simulations produce accurate structures and vibrational spectra for bulk water.\cite{pestana2017initio, marsalek2017quantum, pnas_116_1110} The initial configuration for bulk water was generated using Packmol\cite{martinez2009packmol} and equilibrated for 10 ps using classical revPBE-D3 AIMD simulations in the NVT ensemble with a CSVR thermostat and a 1 fs time step. This was followed by 50 ps  revPBE0-D3 AIMD simulations, with the snapshots from the final 40 ps used for training. 

All 4G-HDCNNP model trainings were performed using n2p2 (see Fig. \ref{fig:workflow}).\cite{singraber2019parallel} The radial and angular symmetry functions for O and H atoms were taken from Ref. \citenum{schran2020committee} and are provided in the Supporting Information Tables S1 and S2. Two hidden layers, each with 15 nodes, were used for both electrostatics and short-range neural networks. The active learning with committee neural network architecture, as implemented in an in-house modified version of AML \cite{schran2020committee}, was employed for efficient sampling and accurate model generations. All active learning parameters were extensively benchmarked. It was found that a committee of four models ($n=4$) is sufficient for all charge, energy, and force trainings (see the Supporting Information Fig. S1). Also, 15 epochs were found to be adequate for both the electrostatics and short-range neural networks. All 4 committee models use the same set of symmetry functions to represent the local environments, but differ in their initial random seed. These variations ensure a sufficiently diverse committee of NNPs. In our active learning workflow, we first selected 20 maximally distant configurations in the first iteration. For all subsequent iterations, 20 configurations with the highest force disagreement among the committee members were selected according to Eq. \ref{eq-9}\cite{schran2020committee}, where $n$ is the committee index and $\alpha$ is the atom index.
\begin{equation}
    \sigma_{F_{\alpha}} = \left[ \frac{1}{n}\sum_{i=1}^{n}\left(\nabla_{\alpha}\Delta E_{i}\right)^{2} \right]^{\frac{1}{2}}
    \label{eq-9}
\end{equation}
Here, $\Delta E_i = E_i - \bar{E}$ is the deviation of the predicted energy of model $i$ from the ensemble mean. The term $\nabla_{\alpha} \Delta E_i$ corresponds to the difference between the force on atom $\alpha$ predicted by model $i$ and the ensemble-averaged force, and $\sigma_{F_{\alpha}}$ represents the force disagreement across the committee. Once the total number of configurations is selected with active learning for a given generation, the final model for that generation is trained with 50 epochs for the electrostatic neural network and 100 epochs for the short-range.

The Generation 1 models were trained on a set of bulk classical configurations with $n=4$ committee models in the QbC cycle. For this purpose, a candidate pool of 4000 distant and random structures was harvested from the revPBE0-D3 AIMD simulation. The selection loop ensured that no configuration was sampled more than once, and evaluated the entire candidate pool at each round. The subsequent generations were trained with both bulk and air-water interface configurations. Once Generation 1 models were trained, MD and PIMD simulations were performed using an in-house modified version of our DL\_POLY Quantum code.\cite{londonDL_POLYQuantumModular2024,london2025dl_poly}. DL\_POLY Quantum is a highly modular, sustainable, and scalable MD simulation platform for large-scale long-time classical and PI simulations of condensed phase and interfacial systems with the essential NQEs included.
Generation 2 models were trained with the configuration pool generated from DL\_POLY Quantum simulations for classical and partially quantum nuclei. At this stage, NQEs were gradually introduced with configurations from partially converged PIMD simulations with 8 beads. Both bulk and interfacial configurations were added to the Generation 2 pool. Similarly, Generation 3 was trained with configurations selected from fully converged PIMD simulations with 32 beads generated from both bulk and air-water interface simulations in DL\_POLY Quantum. Similar to Ref. \citenum{schran2020committee}, a biasing potential was applied during the 4G-HDCNNP MD and PIMD simulations at every stage to stabilize the simulations and prevent the simulations from sampling regions of configuration space with large committee disagreements. Energy, force, and atomic charges for the selected configurations from Generations 2 and 3 were calculated using the same revPBE0-D3 hybrid dispersion-corrected functional and DDEC6 charge partition. All results reported here are obtained using the final Generation 3 models, unless stated otherwise.

\subsection{Simulating IR Spectra of Bulk and Interfacial Water}
%\textit{Simulating IR Spectra of Bulk and Interfacial Water}
Using the trained final Generation 3 models, an initial 50 ps canonical (NVT) sampling was performed for bulk water using the PI Langevin equation (PILE) thermostat\cite{sepCeriotti2010a} with timesteps of 1 fs and 0.5 fs for classical and PIMD simulations, respectively. After discarding equilibration configurations, 10 independent snapshots were extracted for microcanonical (NVE) simulations of 10 ps each with a timestep of 1 fs, 0.5 fs, and 0.25 fs for classical, TRPMD, and PA-CMD simulations, respectively. For each of the NVE trajectories, the initial 2 ps were discarded as equilibration, and the remaining 8 ps trajectory was used for spectral analysis. All dipole moments were calculated using the standard expression for a system of classical point charges: $\boldsymbol{\mu} = \sum_i Q_i \mathbf{R}_i$, where $Q_i$ is the atomic charge associated with atom $i$ and $\mathbf{R}_i$ is its position. Similar to Ref. \citenum{beckmann2022infrared}, every water molecule was first reconstructed with minimum-image connectivity. The center of mass for each water molecule was then shifted to the origin, and the $x$, $y$, and $z$ components of ${\boldsymbol\mu}$ were evaluated. Using these dipoles directly, IR spectra were calculated from the Fourier transform of the total cell dipole-derivative autocorrelation functions with a Hann window cutoff of $\tau_0$ = 0.3 ps. The lineshape sensitivity to different time constants is demonstrated for TRPMD simulations in the Supporting Information Fig. S8. 

%%%%%%%%%%%%%%%%%%%%%%%%%%%%%%%%%%%%%%%%%%%%%
% SECTION 4
%%%%%%%%%%%%%%%%%%%%%%%%%%%%%%%%%%%%%%%%%%%%%
\section{Results and Discussion} \label{sec4}
%\textbf{Results and Discussion}

\subsection{Active Learning and Model Performance}
%\textit{Active Learning and Model Performance}
To benchmark and evaluate our active learning workflow in Generation 1, a separate validation set consisting of 500 distant AIMD-generated bulk configurations was used. For n=2, 4, and 8 models, the charge and force errors on that validation set are illustrated in the Supporting Information Fig. S1. The MAE values converge quickly with increasing the number of committees and plateau within 300 configurations. As such, Generation 1 was trained with 300 QbC selected configurations. Such choices and error levels are consistent with prior reports on water using active learning and 2G-HDNNP committee models, where a few hundred reference configurations were found to be sufficient for achieving high accuracy.\cite{schran2020committee} The charge RMSE for this generation was calculated to be 5.97 me and 5.91 me for the training and test sets, respectively. The energy RMSE is found to be 0.45 meV/atom and 0.41 meV/atom, and the force RMSE was calculated to be 0.05 eV/\AA~for both the training and test sets. These values are consistent with those reported previously (the comparisons are provided in the Supporting Information Table S3).\cite{ko2021fourthgeneration, kocer2024machine} Calculated correlations between the reference and 4G-HDCNNP predicted charges, energies, and forces for the training and test sets are given in the Supporting Information Fig. S2.

\begin{figure}[!h]
    \centering
    \includegraphics[width=0.7\linewidth]{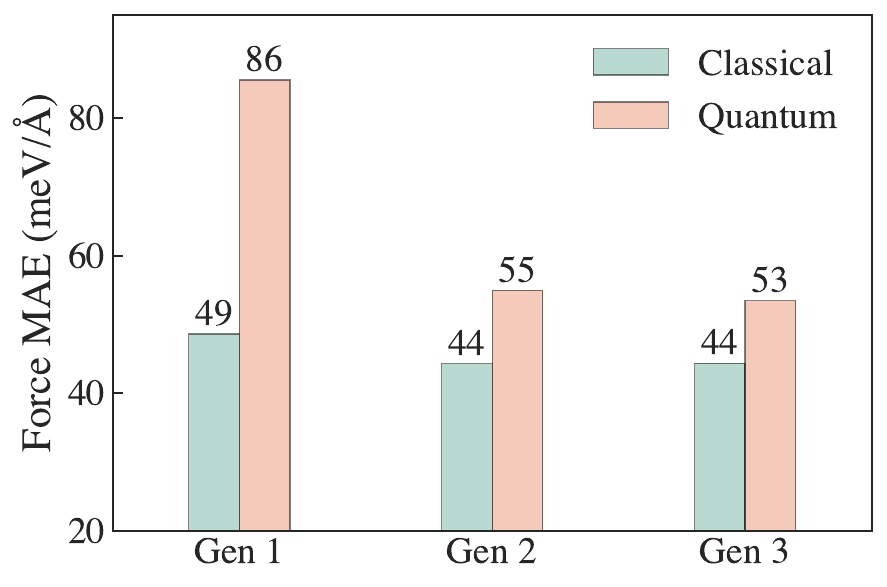}
    \caption{Calculated force mean absolute errors (MAEs) of the 4G-HDCNNP models across all three generations. The MAEe are determined using independently generated validation sets calculated with revPBE0-D3, both for classical and quantum configurations averaged from bulk and interface performances.}
    \label{fig:mae-final}
\end{figure}
Fig. \ref{fig:mae-final} illustrates the calculated MAEs across all three generations. This evaluation was performed on another separate validation set of 2000 configurations from four different trajectories simulated with the final Generation 3 models. The first set of configurations included 500 distinct classical bulk simulation snapshots. The following 500 configurations were generated from a converged 32-bead PIMD simulation of bulk water. The 3rd and 4th sets of 500 configurations each originated from similar classical and PIMD simulations of interfacial water. All reference energy, force, and DDEC6 charges were calculated at the revPBE0-D3 level. As can be seen from Fig. \ref{fig:mae-final}, the classical force MAEs are calculated to be 49 meV/\AA, 44 meV/\AA, and 44 meV/\AA~ for Generation 1-3 models, respectively. For configurations with quantum nuclei, the MAEs are 86 meV/\AA, 55 meV/\AA, and 53 meV/\AA~ for Generation 1-3 models, respectively. Expectedly, the first-generation models trained with only classical configurations perform poorly in describing both bulk and interfacial configurations with quantum nuclei, which improves with subsequent generations as NQEs are introduced during training. All MD and PIMD production simulations using the final trained 4G-HDCNNPs were performed with $n=4$ committee models for Generations 1 and 2. To reduce the computational cost of our simulations without compromising accuracy, for Generation 3, we reduced the number of models to $n=2$, as it was found to yield similar results while being comparatively faster, especially for path integral simulations involving air-water interface systems. This enabled us to perform longer trajectories with no loss of accuracy. The comparison among n = 1, 2, 3, and 4 models, validated using bulk water radial distribution functions (RDFs) calculated with Generation 3 models, is presented in the Supporting Information Fig. S6.

\subsection{Dipole Moment Validation}
\begin{figure}[!t]    
    \centering
    \includegraphics[width=0.65\linewidth]{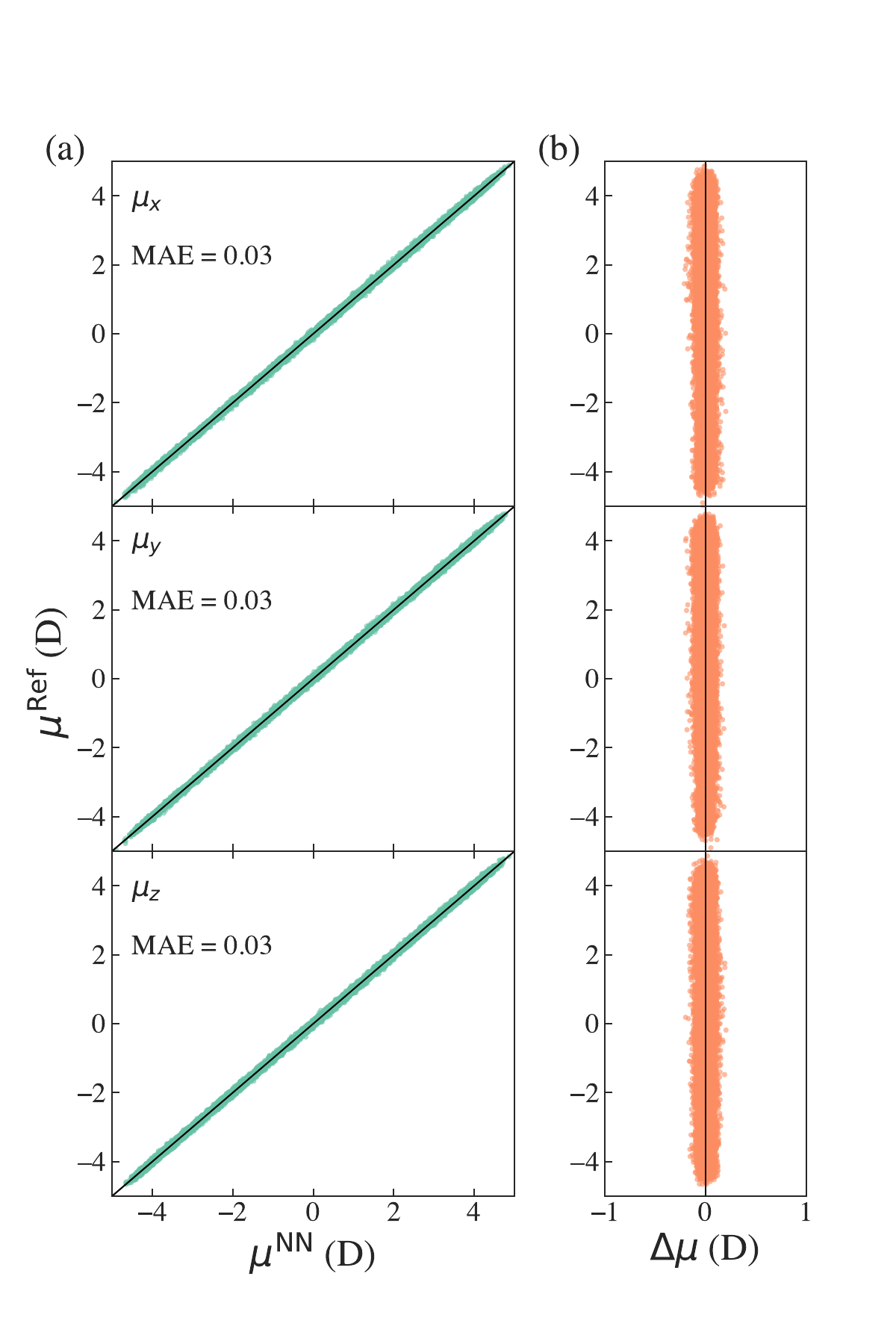}
    \caption{(a) Calculated correlations between the reference (DDEC6) and 4G-HDCNNP predicted dipole moments along $x$, $y$, and $z$ directions. (b) Calculated errors ($\Delta\boldsymbol{\mu} = \boldsymbol{\mu}_{\text{NN}} - \boldsymbol{\mu}_{\text{ref}}$) are also shown.}
    \label{fig:dipole-correlation}
\end{figure}

%\textit{Dipole Moment Validation}
For accurate vibrational spectral simulations, one must also assess electronic observables such as dipole-moment fluctuations in addition to energies and forces. We have therefore adopted dipole moments as part of our evaluation criteria. Fig. \ref{fig:dipole-correlation} illustrates dipole moment correlations for the same classical bulk subset of the 2000 configuration validation set described above. Dipole moment validations for interface and PIMD sets are reported in the Supporting Information Figs. S3-S5. We obtained dipole moment MAEs of 0.03 D for the classical configurations and 0.04 D for the PIMD configurations across both bulk and interfacial validation sets. These validations demonstrate that the 4G-HDCNNP predicted charges can recover the dipole moment to within a few hundredths of a Debye, proving the accuracy of our employed approach in calculating dipoles. Using 4G-HDCNNPs, therefore, provides a robust and practical paradigm for predictive spectral simulations, eliminating the need for explicit training of dipoles, ad-hoc fittings, empirical parameterizations, or charge rescaling to match reference spectra.

\begin{figure*}[!t]
   \centering
   \includegraphics[width=0.99\linewidth]{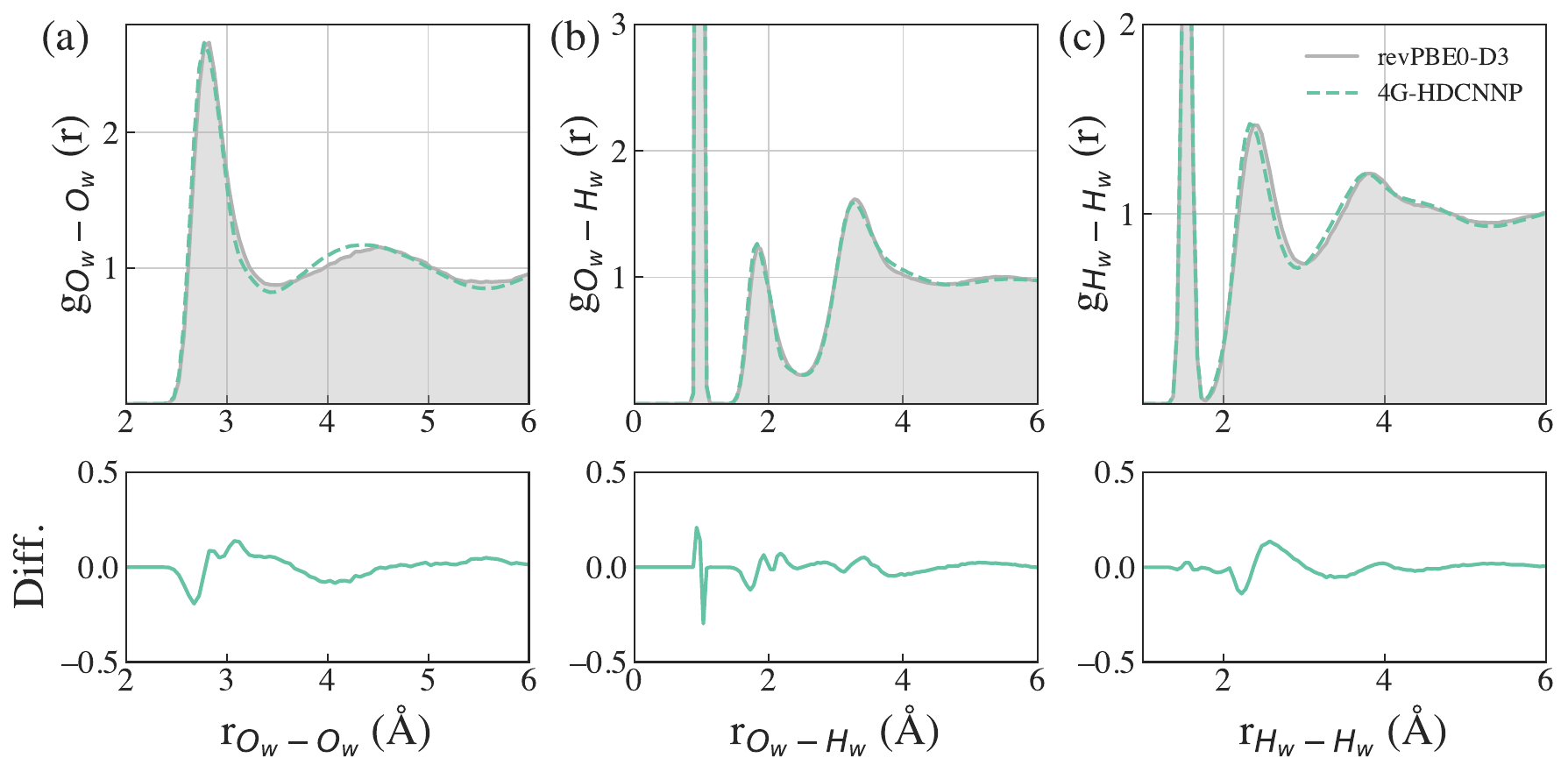}
   \caption{Calculated RDFs for (a) $O_W$--$O_W$ (oxygen of water), (b) $O_W$--$H_W$ (hydrogen of water), and (c) $H_W$--$H_W$ pairs from classical 4G-HDCNNP simulations compared to the reference revPBE0-D3 AIMD data. Calculated errors are shown in the bottom panels. }
   \label{fig:rdf}
\end{figure*}

Fig. \ref{fig:rdf} compares the calculated RDFs for bulk water using the final Generation 3 models compared to reference revPBE0-D3. 4G-HDCNNP-generated RDFs from classical MD simulations are in excellent agreement with the reference classical AIMD calculated RDFs for $O_W$--$O_W$, $O_W$--$H_W$, and $H_W$--$H_W$ pairs. This demonstrates that our trained 4G-HDCNNPs can accurately capture the dynamic H-bond network of water at finite temperatures. To evaluate finite-size effects, classical MD simulations were also performed for a larger box of 17.53~\AA~ containing 180 water molecules, corresponding to the same density of 0.997 g/cm$^{3}$ at room temperature, where negligible differences were found (see the Supporting Information Fig. S7).

\begin{figure}[!t]
    \centering
    \includegraphics[width=0.7\linewidth]{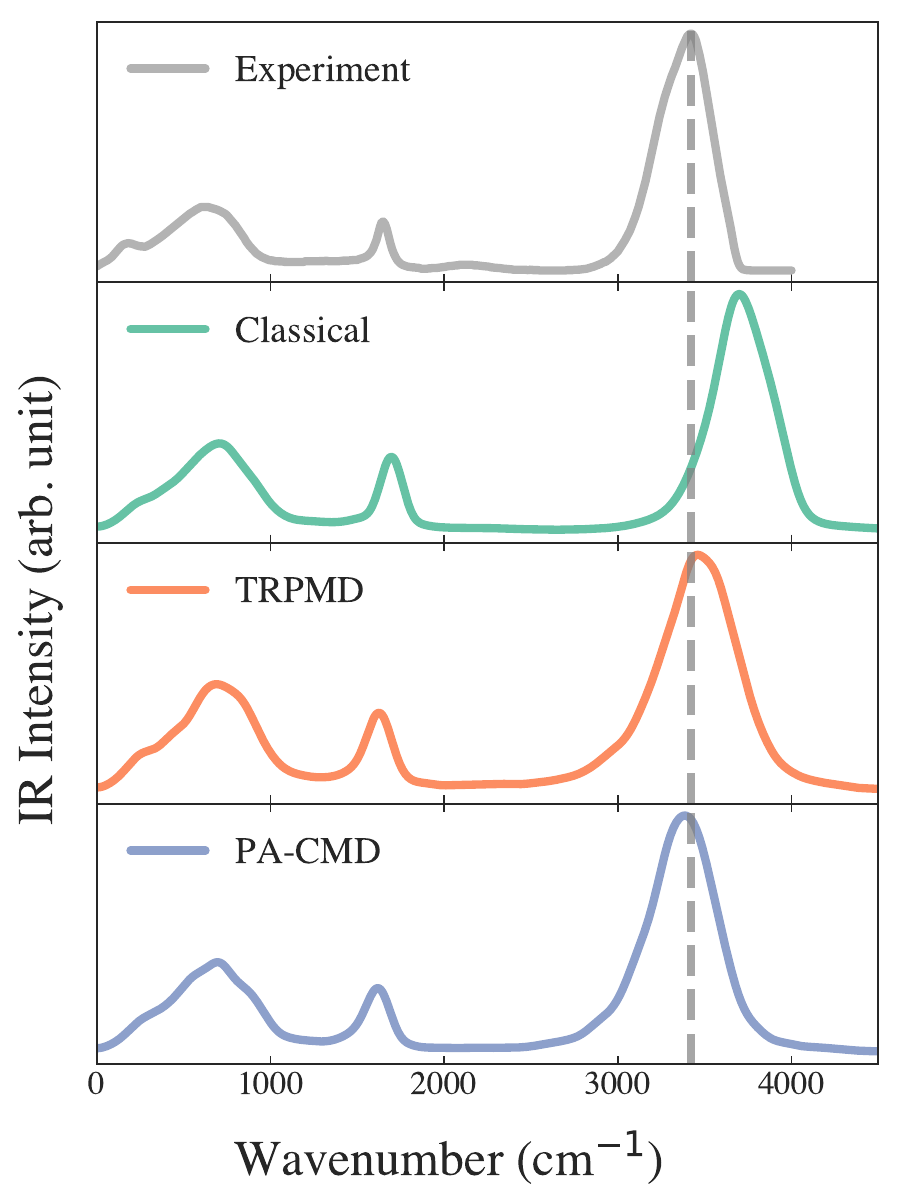}
    \caption{4G-HDCNNP calculated MD, TRPMD, and PA-CMD simulated IR spectra of bulk water compared to the experiment.\cite{augBertie1996} The dashed line represents the OH-stretch maximum peak position from the experiment. }
    \label{fig:ir-bulk}
\end{figure}

\begin{figure}[t]
    \centering
    \includegraphics[width=0.7\linewidth]{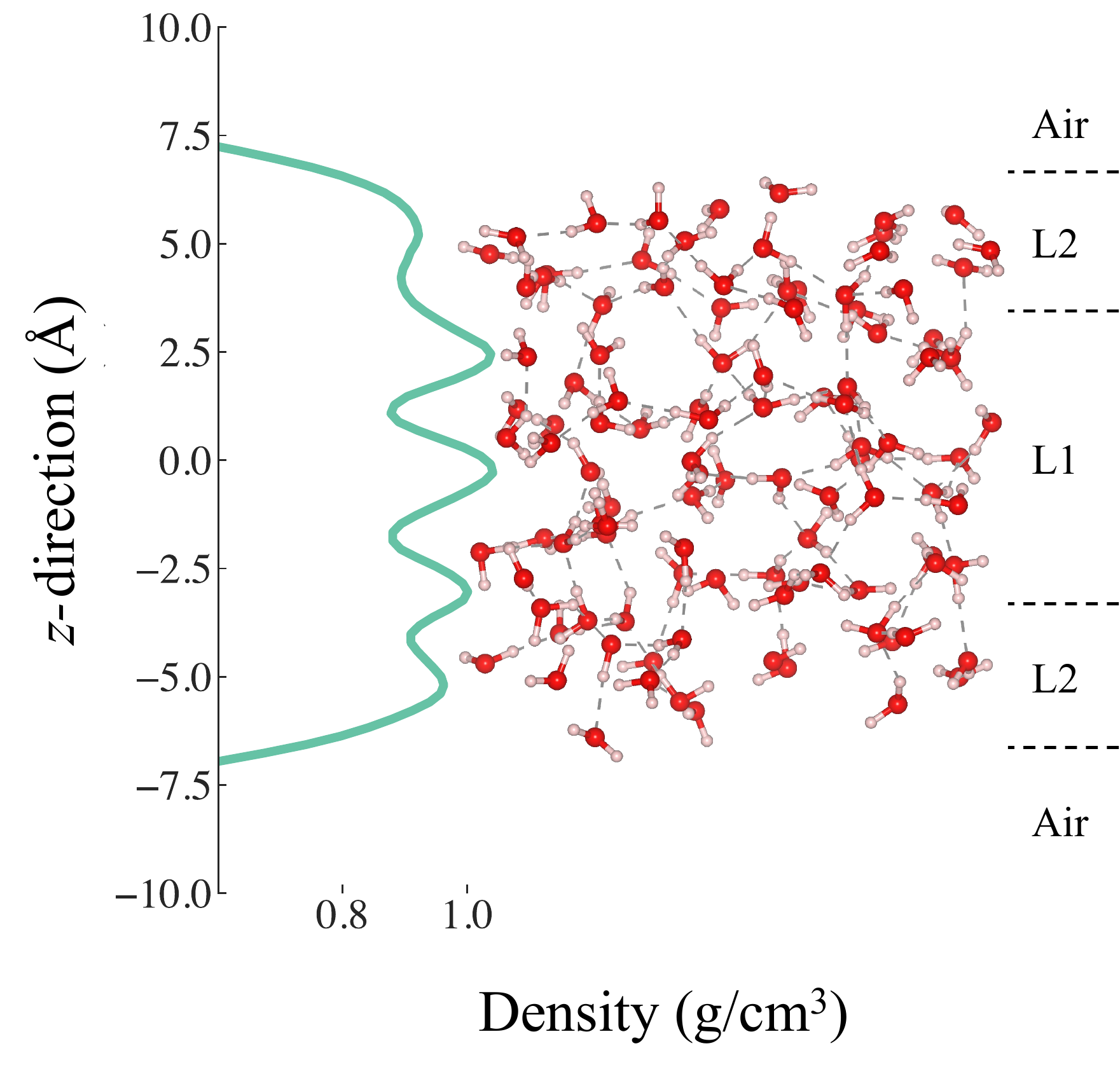}
    \caption{Calculated density profile of water in the $z$ direction for the air-water interface system, along with the two water layers considered in this work.}
    \label{fig:density}
\end{figure}

\begin{figure}[t]
    \centering
    \includegraphics[width=0.6\linewidth]{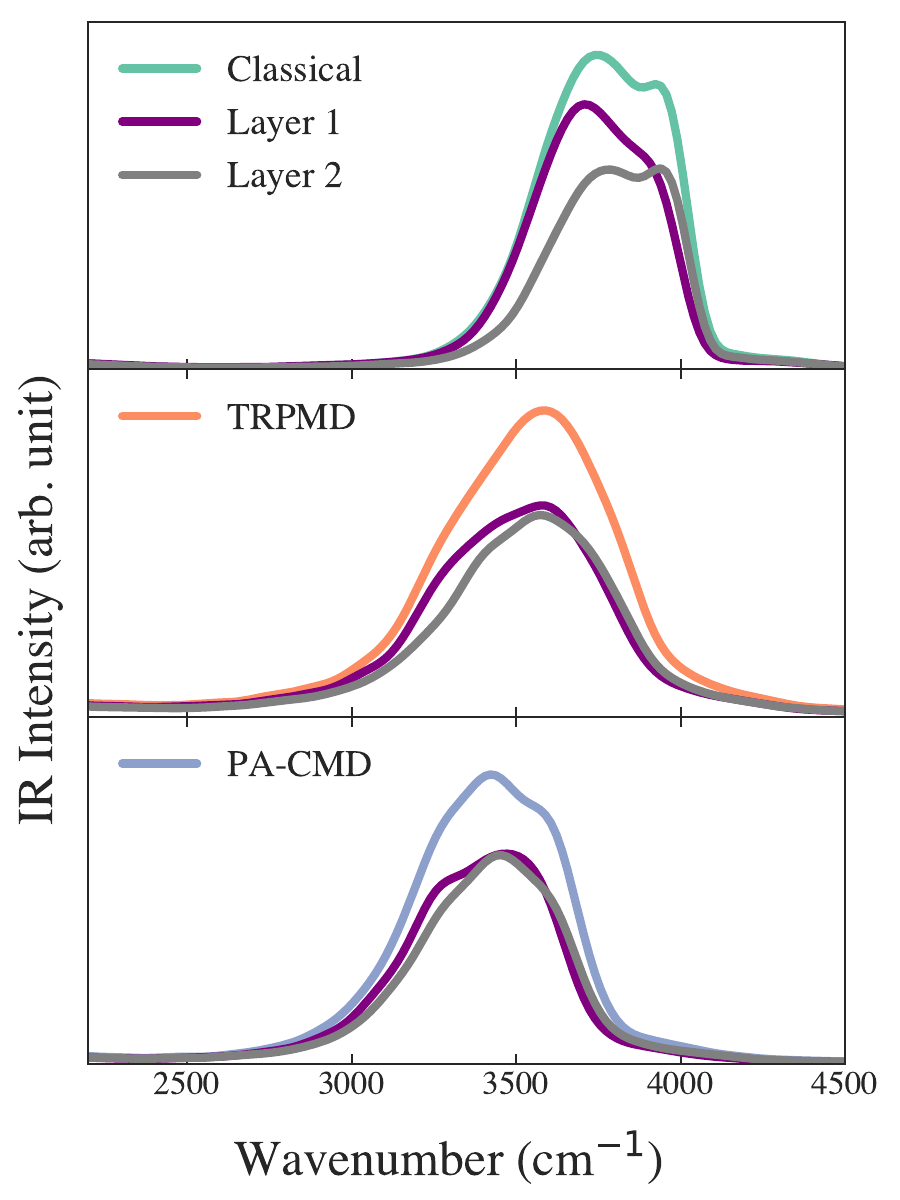}
    \caption{4G-HDCNNP simulated IR spectra of the air-water interface system. Layer 1 corresponds to the bulk-like water at the center of the box, while Layer 2 denotes the surface layer. The deconvoluted spectra are given in the Supporting Information Fig. S9.}
    \label{fig:ir-interface}
\end{figure}

\subsection{IR Spectra of Bulk Water}
%\textit{IR Spectra of Bulk Water}
The simulated IR spectra of liquid bulk water at 298 K obtained from classical, PA-CMD, and TRPMD simulations using the final trained 4G-HDCNNPs are presented in Fig. \ref{fig:ir-bulk}.  The experimental IR peak for the O-H stretch region appears at 3420 cm$^{-1}$. In classical simulations, the O–H stretch exhibits a pronounced blue shift of 274 cm$^{-1}$ relative to the experiment due to the neglect of NQEs. The blue shift for the classical spectra also appears in the bending region, but to a much lesser extent. The inclusion of NQEs through real-time path integral methods, TRPMD and PA-CMD, largely corrects this deviation. TRPMD generates a stretch peak with a modest blue shift of 24 cm$^{-1}$ and small broadening effects. The PA-CMD produces a stretch maximum with a slight red shift of 26 cm$^{-1}$. The artificial broadening and red shift of the O-H stretch peak of bulk water at room temperature, as obtained from TRPMD and PA-CMD, respectively, are in agreement with the known problems associated with these methods, as summarized in our previous works.\cite{limbu2025hcmd, london2025dl_poly,londonDL_POLYQuantumModular2024} All three methods more or less agree on the positions and relative intensities of the peaks for the librational mode, indicating that the 4G models accurately capture the dipole derivatives responsible for low-frequency absorption and can reliably predict underlying vibrational physics. 

\subsection{IR Spectra of the Air-Water Interface}
%\textit{IR Spectra of the Air-Water Interface}
For the IR spectra of the air-water interface, using the density profile of water along the $z$-direction, the simulation box was partitioned into two different layers, namely Layers 1 and 2 (Fig. \ref{fig:density}). Layer 1 represents the bulk-like water, which sits close to the center of the box ($z=0$). The thickness of the bulk-like water layer was kept as $\approx$8.00~\AA. Layer 2 on each side represents the interface layers with a thickness of $\approx$4.00~\AA~each. The water density is highest and bulk-like near the center, and shows a gradual depletion toward the vapor phase.
Our Layer-resolved IR spectra calculated using classical, TRPMD, and PA-CMD methods for the air-water interface at 298 K are presented in Fig. \ref{fig:ir-interface} (the deconvoluted spectra are provided in the Supporting Information Fig. S9). For each method, the total spectra are accompanied by the spectra from Layers 1 and 2 corresponding to bulk-like and interface-like water molecules, respectively. In the classical simulations, the total spectra exhibit two well-separated peaks. The bulk-like band centered around 3700 cm$^{-1}$ is dominated by the Layer-1 water molecules, which reside at the center of the box (Fig. \ref{fig:density}). The high frequency peak around 3944 cm$^{-1}$ arises from Layer-2, which is mostly comprised of interfacial water molecules with abundant single H-bond donors and dangling free OH bonds. Similar to the bulk spectra, all classical peaks are blue-shifted due to the neglect of the NQEs. The O-H stretch peaks are red-shifted once the NQEs are added through TRPMD and PA-CMD. The TRPMD peak is artificially broadened, which hides the distinct doublet feature characteristic of interfacial systems. The PA-CMD spectrum is expectedly red-shifted and slightly broadened, though to a lesser extent, compared to TRPMD due to its well-known curvature problem. However, the total PA-CMD peak still exhibits a doublet feature, although not as pronounced as the classical spectrum.

%%%%%%%%%%%%%%%%%%%%%%%%%%%%%%%%%%%%%%%%%%%%%
% CONCLUSION
%%%%%%%%%%%%%%%%%%%%%%%%%%%%%%%%%%%%%%%%%%%%%
\section{Conclusions} \label{sec5}
%\textbf{Conclusions\newline}

In summary, this work demonstrates that 4G-HDCNNPs efficiently trained using active learning and query-by-committee can be utilized for predictive quantum vibrational spectral simulations without the need for explicit training of dipole moments, ad hoc fitting, or parameterizations. This feat is achieved by combining non-local effects from 4G-HDCNNPs with nuclear quantum effects from Feynman's imaginary-time path integral formalism. Analyses of IR spectra for bulk and air-water interface test cases demonstrate that the 4G-HDCNNP predicted charges are reliable for dipole and IR spectral calculations. The adapted methodology in this work is general and applicable to other complex systems and environments with significant long-range charge transfer effects.
Future works will focus on (i) calculating spectra through combining 4G-HDCNNPs with recently introduced curvature-free CMD-derived real-time methods, including Te-PIGS,\cite{tepigs} f-QCMD,\cite{decFletcher2021,octLawrence2023} and our very own h-CMD\cite{limbu2025hcmd} methods, (ii) applying 4G-HDCNNPs to complex reactive electrochemical environments at different condensed phases and interfaces dominated by non-local charge transfer interactions, and (iii) including constant potentials for the accurate simulations of different dynamical and spectral properties in the presence of external electric fields. We hope that these developments will facilitate the accurate calculation of quantum dynamical properties and vibrational spectra, addressing important and long-standing questions at the heart of energy conversion and storage processes, among others.

\begin{suppinfo}
The data that support the findings of this study are provided in the main text and the accompanying Supporting Information. Additional data are available from the corresponding author upon request.
\end{suppinfo}

\begin{acknowledgement}
Simulations presented in this work used resources from Bridges-2 at Pittsburgh Supercomputing Center through allocations PHY230030P and CHE250024P from the Advanced Cyberinfrastructure Coordination Ecosystem: Services \& Support (ACCESS) program,\cite{access} which is supported by National Science Foundation grants \#2138259, \#2138286, \#2138307, \#2137603, and \#2138296. The authors also acknowledge the HPC center at UMKC for providing computing resources and support.
\end{acknowledgement}

\bibliography{manuscript}

\begin{tocentry}
\includegraphics[width=0.99\linewidth]{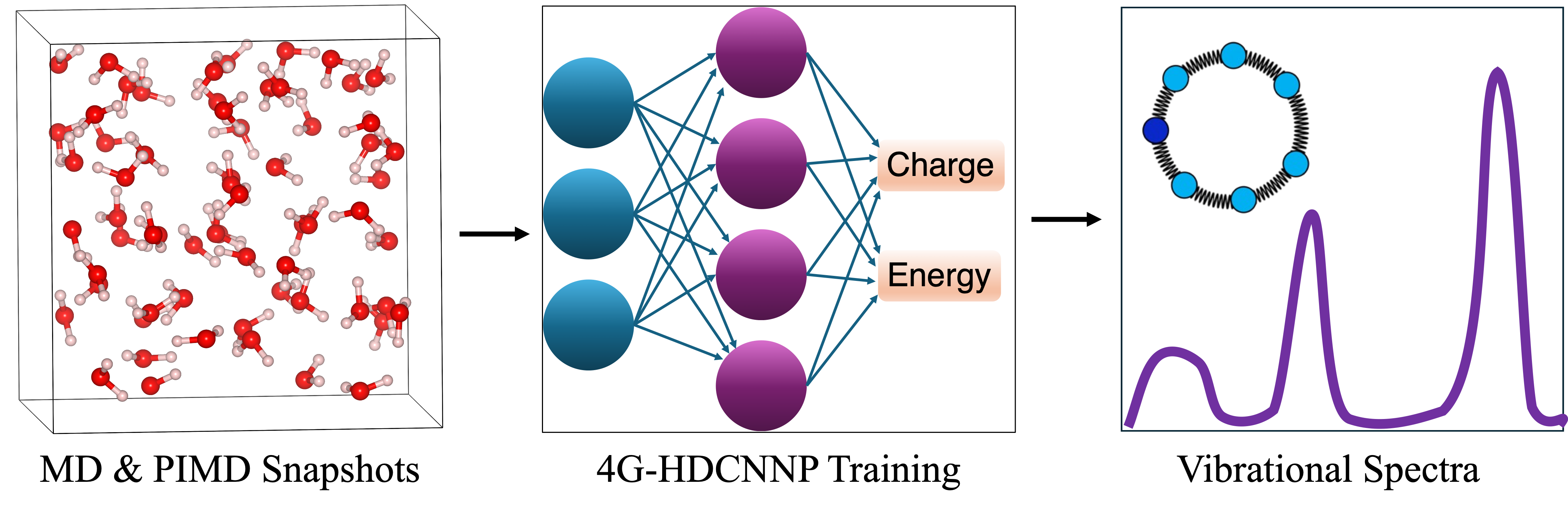}
TOC entry
\end{tocentry}

\end{document}

\newpage
\tableofcontents
\newpage

\begin{table}[!h]
    \addcontentsline{toc}{table}{Table \ref{tab:sf_radial}. Parameters for the radial symmetry functions.}
    \centering
    \begin{tabular}{cccccc}
        \hline
        ID & Atom & Neighbor & $\eta$ (1/Bohr$^2$) & $R_s$ (Bohr) & $R_c$ (Bohr) \\
        \hline
        1  & H & H & 0.001 & 0.0 & 12 \\
        2  & H & H & 0.010 & 0.0 & 12 \\
        3  & H & H & 0.030 & 0.0 & 12 \\
        4  & H & H & 0.060 & 0.0 & 12 \\
        5  & H & H & 0.150 & 1.9 & 12 \\
        6  & H & H & 0.300 & 1.9 & 12 \\
        7  & H & H & 0.600 & 1.9 & 12 \\
        8  & H & H & 1.500 & 1.9 & 12 \\
        9  & H & O & 0.001 & 0.0 & 12 \\
        10 & H & O & 0.010 & 0.0 & 12 \\
        11 & H & O & 0.030 & 0.0 & 12 \\
        12 & H & O & 0.060 & 0.0 & 12 \\
        13 & H & O & 0.150 & 0.9 & 12 \\
        14 & H & O & 0.300 & 0.9 & 12 \\
        15 & H & O & 0.600 & 0.9 & 12 \\
        16 & H & O & 1.500 & 0.9 & 12 \\
        17 & O & H & 0.001 & 0.0 & 12 \\
        18 & O & H & 0.010 & 0.0 & 12 \\
        19 & O & H & 0.030 & 0.0 & 12 \\
        20 & O & H & 0.060 & 0.0 & 12 \\
        21 & O & H & 0.150 & 0.9 & 12 \\
        22 & O & H & 0.300 & 0.9 & 12 \\
        23 & O & H & 0.600 & 0.9 & 12 \\
        24 & O & H & 1.500 & 0.9 & 12 \\
        25 & O & O & 0.001 & 0.0 & 12 \\
        26 & O & O & 0.010 & 0.0 & 12 \\
        27 & O & O & 0.030 & 0.0 & 12 \\
        28 & O & O & 0.060 & 0.0 & 12 \\
        29 & O & O & 0.150 & 4.0 & 12 \\
        30 & O & O & 0.300 & 4.0 & 12 \\
        31 & O & O & 0.600 & 4.0 & 12 \\
        32 & O & O & 1.500 & 4.0 & 12 \\
        \hline
    \end{tabular}
    \caption{Parameters for the radial symmetry functions.}
    \label{tab:sf_radial}
\end{table}

\clearpage
\begin{table}[!h]
    \addcontentsline{toc}{table}{Table \ref{tab:sf_angular}. Parameters for the angular symmetry functions.}
    \centering
    \begin{tabular}{ccccccccc}
        \hline
        ID & Atom & Neighbor 1 & Neighbor 2 & $\eta$ (1/Bohr$^2$)  & $\lambda$ & $\zeta$ & $R_c$ (Bohr) \\
        \hline
        1 & H & O & H & 0.200 & 1.0 & 1.0 & 12 \\
        2 & O & H & H & 0.070 & 1.0 & 1.0 & 12 \\
        3 & H & O & H & 0.070 & 1.0 & 1.0 & 12 \\
        4 & O & H & H & 0.070 & -1.0 & 1.0 & 12 \\
        5 & H & O & H & 0.070 & -1.0 & 1.0 & 12 \\
        6 & O & H & H & 0.030 & 1.0 & 1.0 & 12 \\
        7 & H & O & H & 0.030 & 1.0 & 1.0 & 12 \\
        8 & O & H & H & 0.030 & -1.0 & 1.0 & 12 \\
        9 & H & O & H & 0.030 & -1.0 & 1.0 & 12 \\
        10 & O & H & H & 0.010 & 1.0 & 4.0 & 12 \\
        11 & H & O & H & 0.010 & 1.0 & 4.0 & 12 \\
        12 & O & H & H & 0.010 & -1.0 & 4.0 & 12 \\
        13 & H & O & H & 0.010 & -1.0 & 4.0 & 12 \\
        14 & O & O & H & 0.030 & 1.0 & 1.0 & 12 \\
        15 & O & O & H & 0.030 & -1.0 & 1.0 & 12 \\
        16 & O & O & H & 0.001 & 1.0 & 4.0 & 12 \\
        17 & O & O & H & 0.001 & -1.0 & 4.0 & 12 \\
        18 & H & O & O & 0.030 & 1.0 & 1.0 & 12 \\
        19 & H & O & O & 0.030 & -1.0 & 1.0 & 12 \\
        20 & H & O & O & 0.001 & 1.0 & 4.0 & 12 \\
        21 & H & O & O & 0.001 & -1.0 & 4.0 & 12 \\
        22 & O & O & O & 0.030 & 1.0 & 1.0 & 12 \\
        23 & O & O & O & 0.030 & -1.0 & 1.0 & 12 \\
        24 & O & O & O & 0.001 & 1.0 & 4.0 & 12 \\
        25 & O & O & O & 0.001 & -1.0 & 4.0 & 12 \\
        \hline
    \end{tabular}
    \caption{Parameters for the angular symmetry functions.}
    \label{tab:sf_angular}
\end{table}

\clearpage

\begin{figure*}[!h]
    \addcontentsline{toc}{figure}{Figure \ref{fig:al-benchmark}. Active learning benchmark.}
   \centering
   \includegraphics[width=0.99\linewidth]{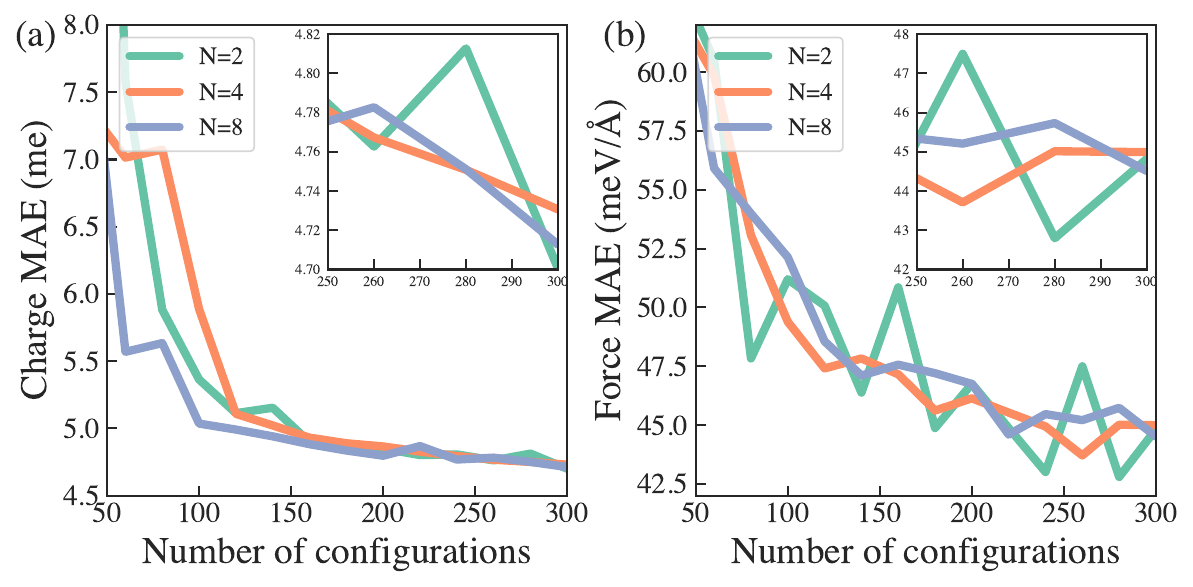}
   \caption{Active learning benchmark: (a) Charge MAE and (b) Force MAE for the number of models ($n$) in each committee.}
   \label{fig:al-benchmark}
\end{figure*}

\clearpage
\begin{figure*}[!ht]    
    \addcontentsline{toc}{figure}{Figure \ref{fig:correlation}. Calculated correlation between the reference and 4G-HDCNNP-predicted charges, energies, and forces.}
    \centering
    \includegraphics[width=0.7\linewidth]{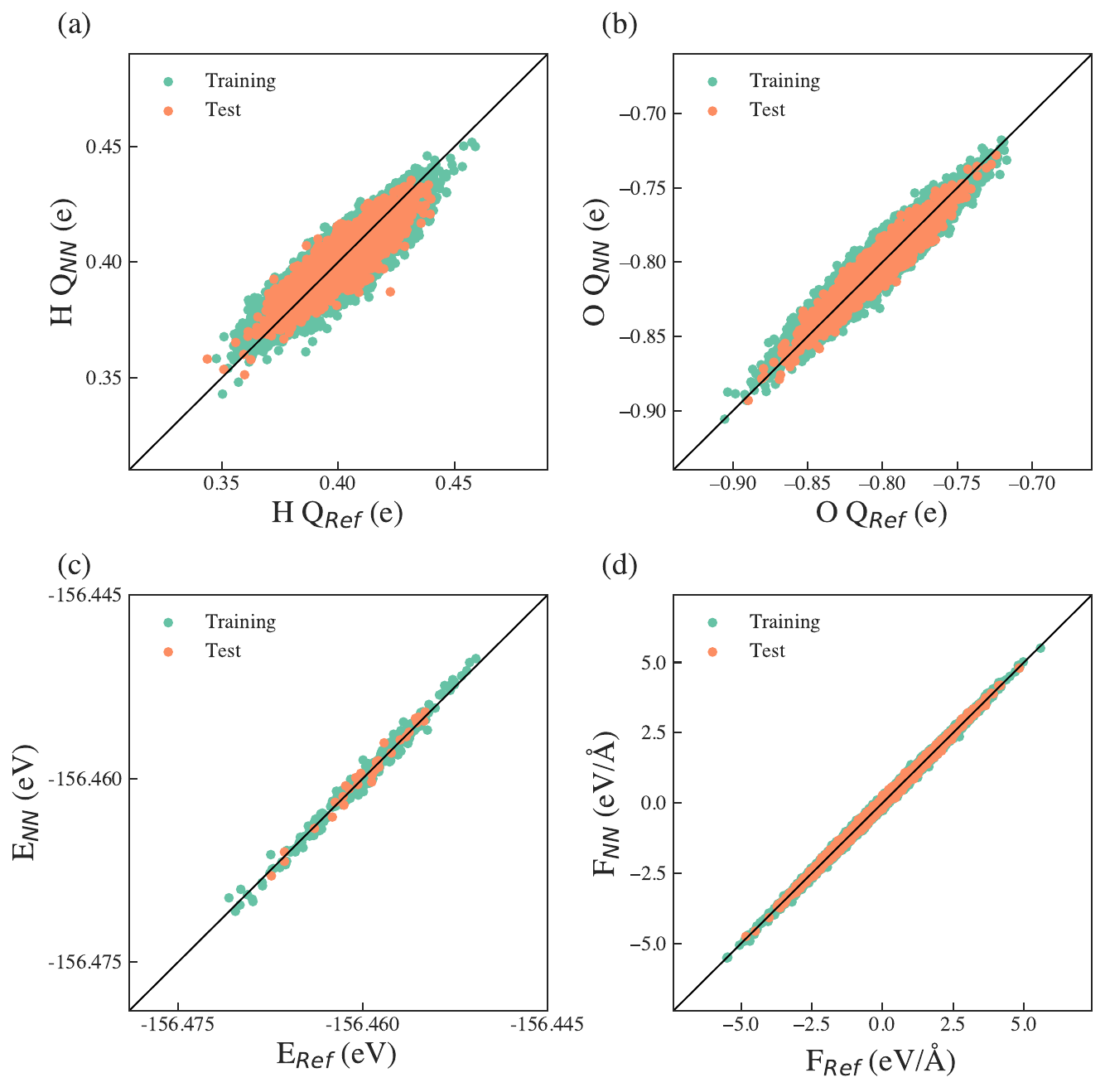}
    \caption{Calculated correlation between the reference and 4G-HDCNNP-predicted (a,b) charges, (c) energies, and (d) forces for both training and test sets.}
    \label{fig:correlation}
\end{figure*}

\clearpage
\begin{table}[!h]
    \addcontentsline{toc}{table}{Table \ref{table:rmse}. Comparison of model errors with the literature.}
     \centering
     \renewcommand{\arraystretch}{1.2}
     \begin{tabular}{lcc}
         \toprule
         & \textbf{Train RMSE} & \textbf{Test RMSE} \\
         \midrule
         \multicolumn{3}{l}{\textbf{Q (me)}} \\
         This work & 5.97 & 5.91 \\        \citeauthor{kocer2024machine}\cite{kocer2024machine} & 2.90 & 2.90 \\        \citeauthor{ko2021fourthgeneration}\cite{ko2021fourthgeneration} & 15.87 & 15.83 \\
         \midrule
         \multicolumn{3}{l}{\textbf{E (meV/atom)}} \\
         This work & 0.45 & 0.41 \\        \citeauthor{kocer2024machine}\cite{kocer2024machine} & 0.20 & 0.21 \\        \citeauthor{ko2021fourthgeneration}\cite{ko2021fourthgeneration} & 0.47 & 0.48 \\        \citeauthor{schran2020committee}\cite{schran2020committee} (2G) & 0.76 & - \\
         \midrule
         \multicolumn{3}{l}{\textbf{F (eV/\AA)}} \\
         This work & 0.05 & 0.05 \\        \citeauthor{kocer2024machine}\cite{kocer2024machine} & 0.06 & 0.06 \\        \citeauthor{ko2021fourthgeneration}\cite{ko2021fourthgeneration} & 0.03 & 0.03 \\        \citeauthor{schran2020committee}\cite{schran2020committee} (2G) & 0.04 & - \\
         \bottomrule
     \end{tabular}
     \caption{Calculated RMSEs for the energy (meV/atom), force (eV/\AA), and charges (me) of both the training and test sets of the committee 4G-HDNNPs developed in this work compared to data from the literature.}
     \label{table:rmse}
         \addcontentsline{toc}{Table}{Table \ref{table:rmse}. Comparison of RMSEs from this work to the literature.}
\end{table}

\clearpage
\begin{figure*}[!ht]    
    \addcontentsline{toc}{figure}{Figure \ref{fig:dipole-bulk-32}. Calculated correlation between the reference and final model predicted dipole moment for the bulk PIMD validation set.}
    \centering
    \includegraphics[width=0.7\linewidth]{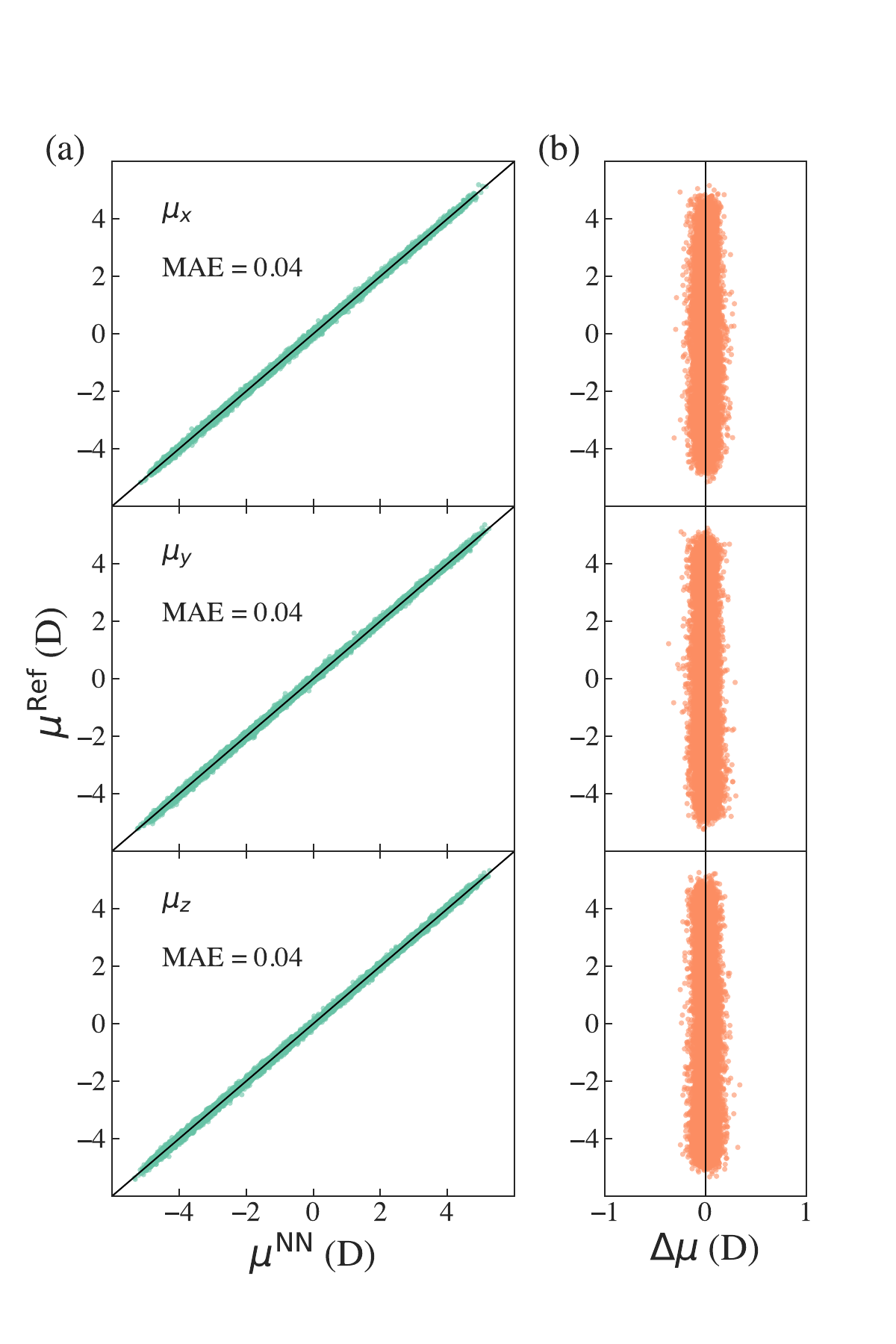}
    \caption{Calculated correlation between the reference and final model predicted dipole moment of bulk PIMD validation set, similar to Fig. 3 in the main text.}
    \label{fig:dipole-bulk-32}
\end{figure*}

\begin{figure*}[!ht]    
    \addcontentsline{toc}{figure}{Figure \ref{fig:dipole-interface-1}. Calculated correlation between the reference and final model predicted dipole moment of the air-water interface classical validation set.}
    \centering
    \includegraphics[width=0.7\linewidth]{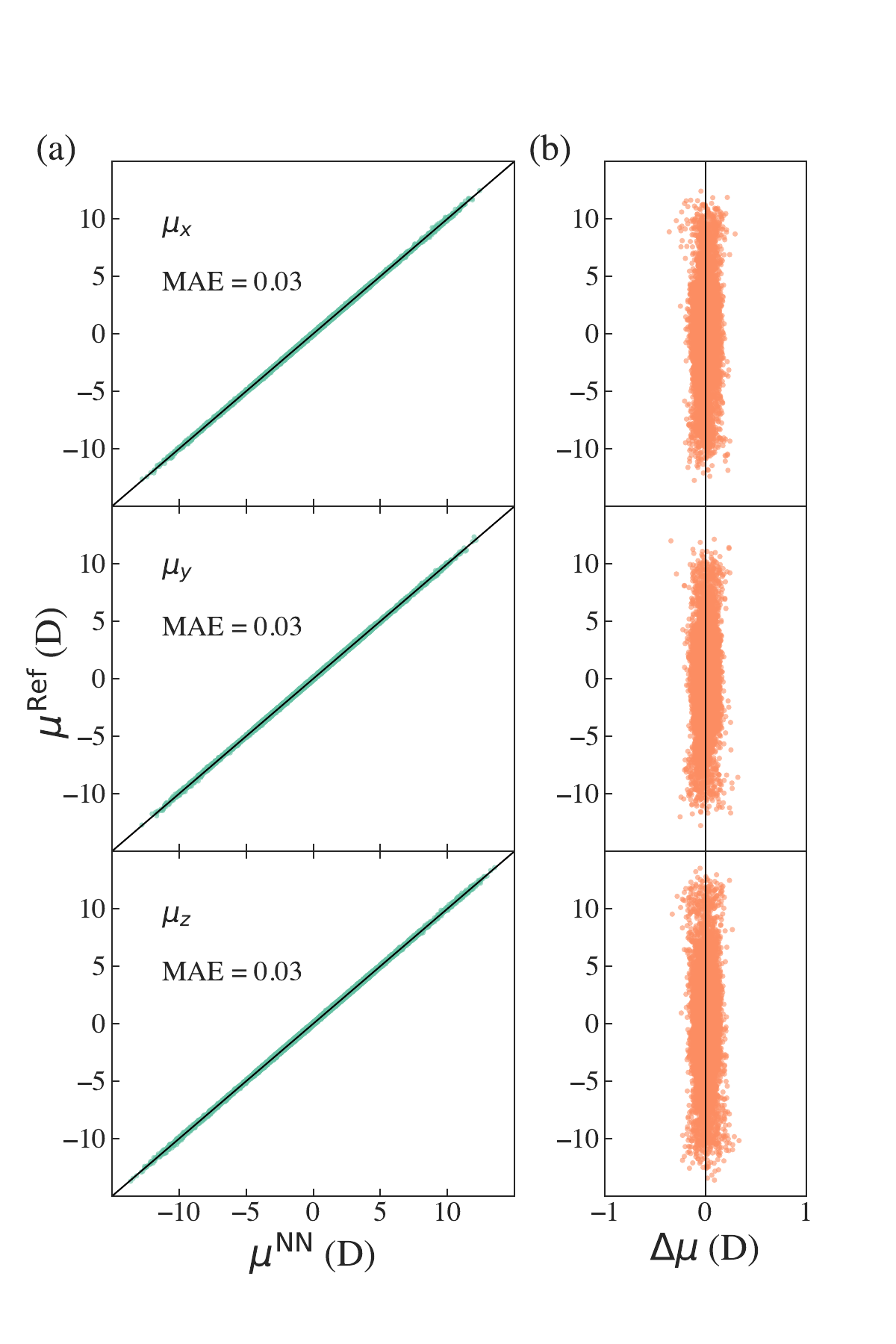}
    \caption{Calculated correlation between the reference and final model predicted dipole moment of the air-water interface classical validation set, similar to Fig. 3 in the main text.}
    \label{fig:dipole-interface-1}
\end{figure*}

\begin{figure*}[!ht]    
    \addcontentsline{toc}{figure}{Figure \ref{fig:dipole-interface-32}. Calculated correlation between the reference and final model predicted dipole moment of the air-water interface PIMD validation set.}
    \centering
    \includegraphics[width=0.7\linewidth]{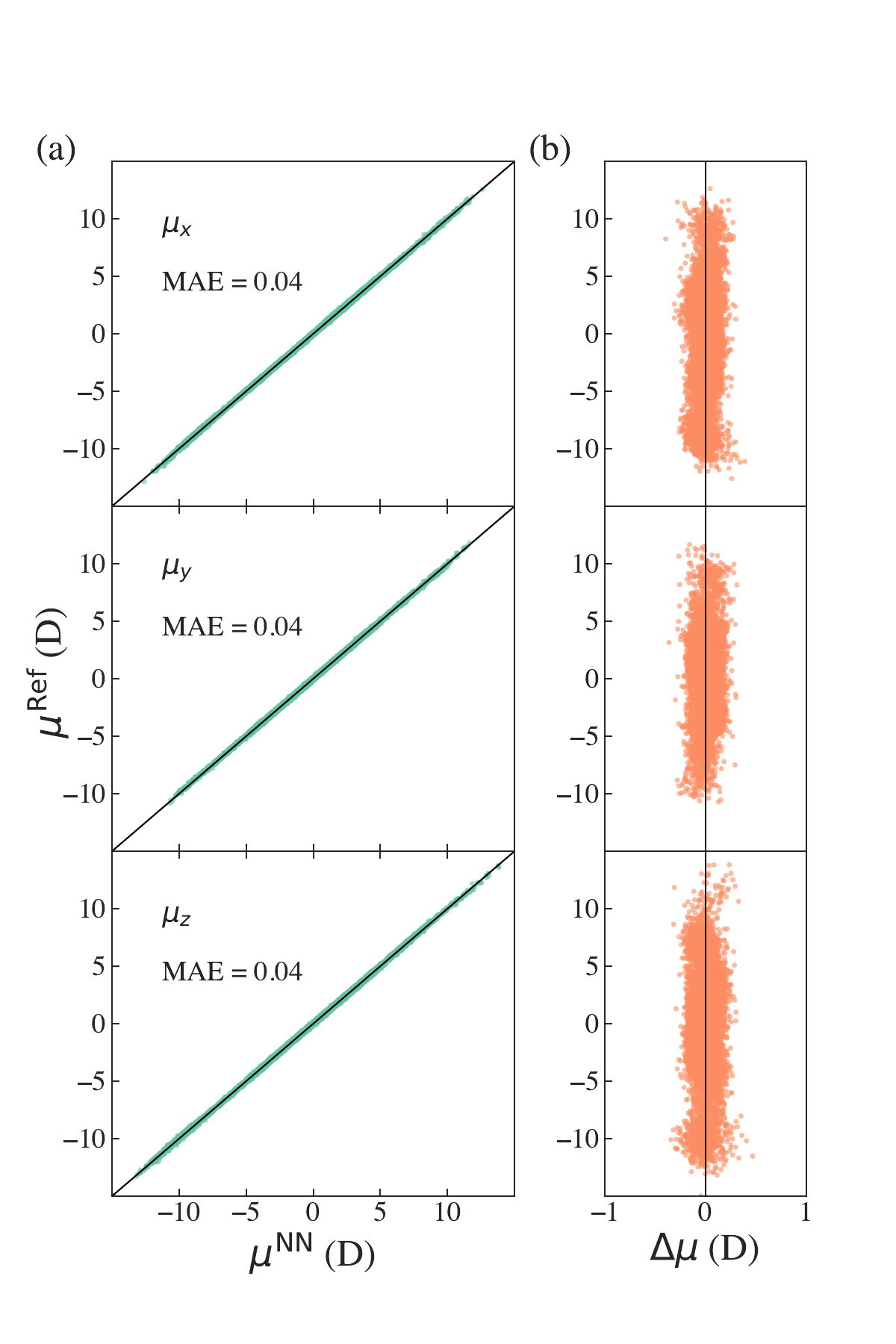}
    \caption{Calculated correlation between the reference and final model predicted dipole moment of the air-water interface PIMD validation set, similar to Fig. 3 in the main text.}
    \label{fig:dipole-interface-32}
\end{figure*}

% \clearpage
% \begin{figure*}[!h]
%     \addcontentsline{toc}{figure}{Figure \ref{fig:convergence}. Temperature, energy, and charge convergence plots from 4G-HDCNNP MD simulations.}
%    \centering
%    \includegraphics[width=0.99\linewidth]{../figs/si-convergence}
%    \caption{(a) Temperature, (b) energy, and (c)  charge convergence plots from 4G-HDCNNP MD simulations.}
%    \label{fig:convergence}
% \end{figure*}

\clearpage
\begin{figure*}[!h]
    \addcontentsline{toc}{figure}{Figure \ref{fig:model_compare}. RDF comparisons between $n=1-4$ committee models from classical MD simulations.}
   \centering
   \includegraphics[width=0.99\linewidth]{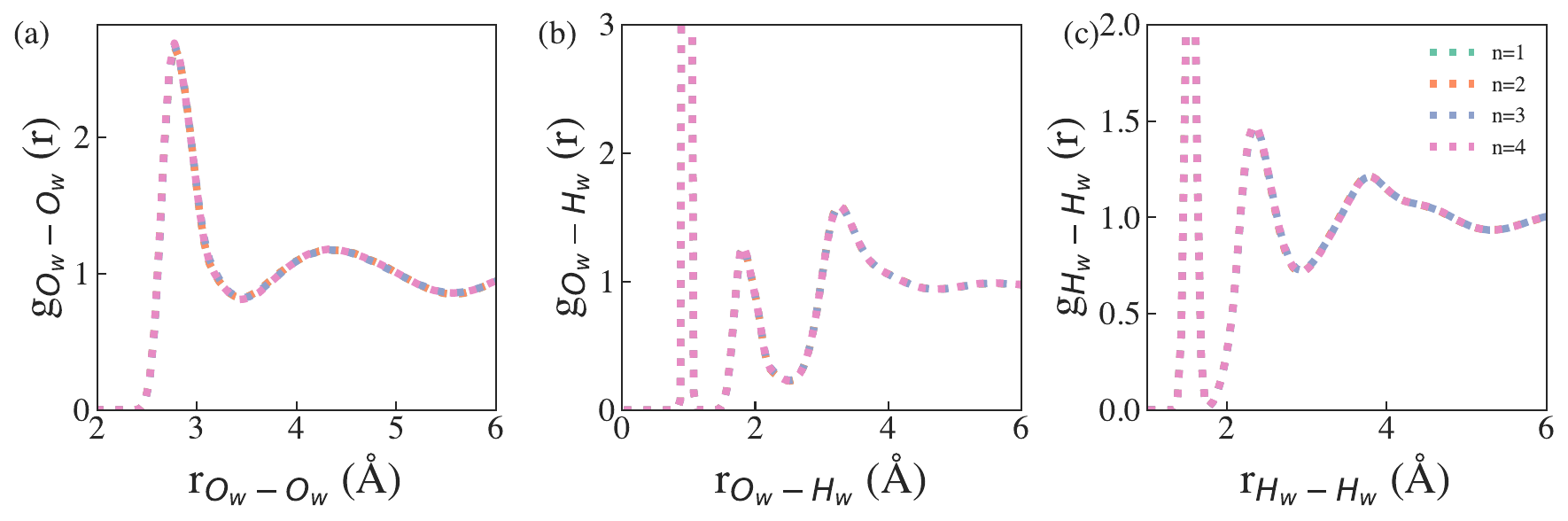}
   \caption{RDF comparisons between $n=1-4$ committee models from classical MD simulations.}
   \label{fig:model_compare}
\end{figure*}

\clearpage
\begin{figure*}[!h]
    \addcontentsline{toc}{figure}{Figure \ref{fig:finite-size}. RDF comparisons between simulation boxes with different numbers of bulk water molecules from classical MD simulations.}
   \centering
   \includegraphics[width=0.99\linewidth]{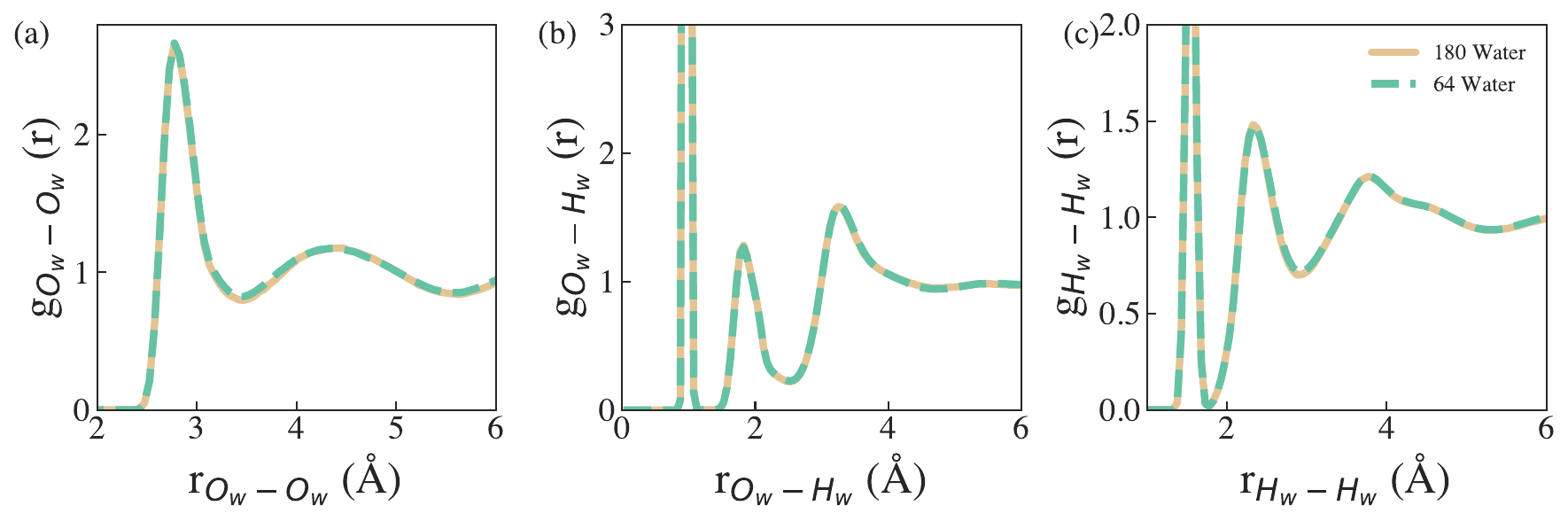}
   \caption{Classical 4G-HDCNNP-calculated RDFs for (a) $O_W$--$O_W$ (oxygen of water), (b) $O_W$--$H_W$ (hydrogen of water), and (c) $H_W$--$H_W$ pairs for simulation boxes containing 180 and 64 water molecules.}
   \label{fig:finite-size}
\end{figure*}

\clearpage
\begin{figure*}[!h]
    \addcontentsline{toc}{figure}{Figure \ref{fig:tau_compare}. Comparison of Hann window cutoff.}
   \centering
   \includegraphics[width=0.7\linewidth]{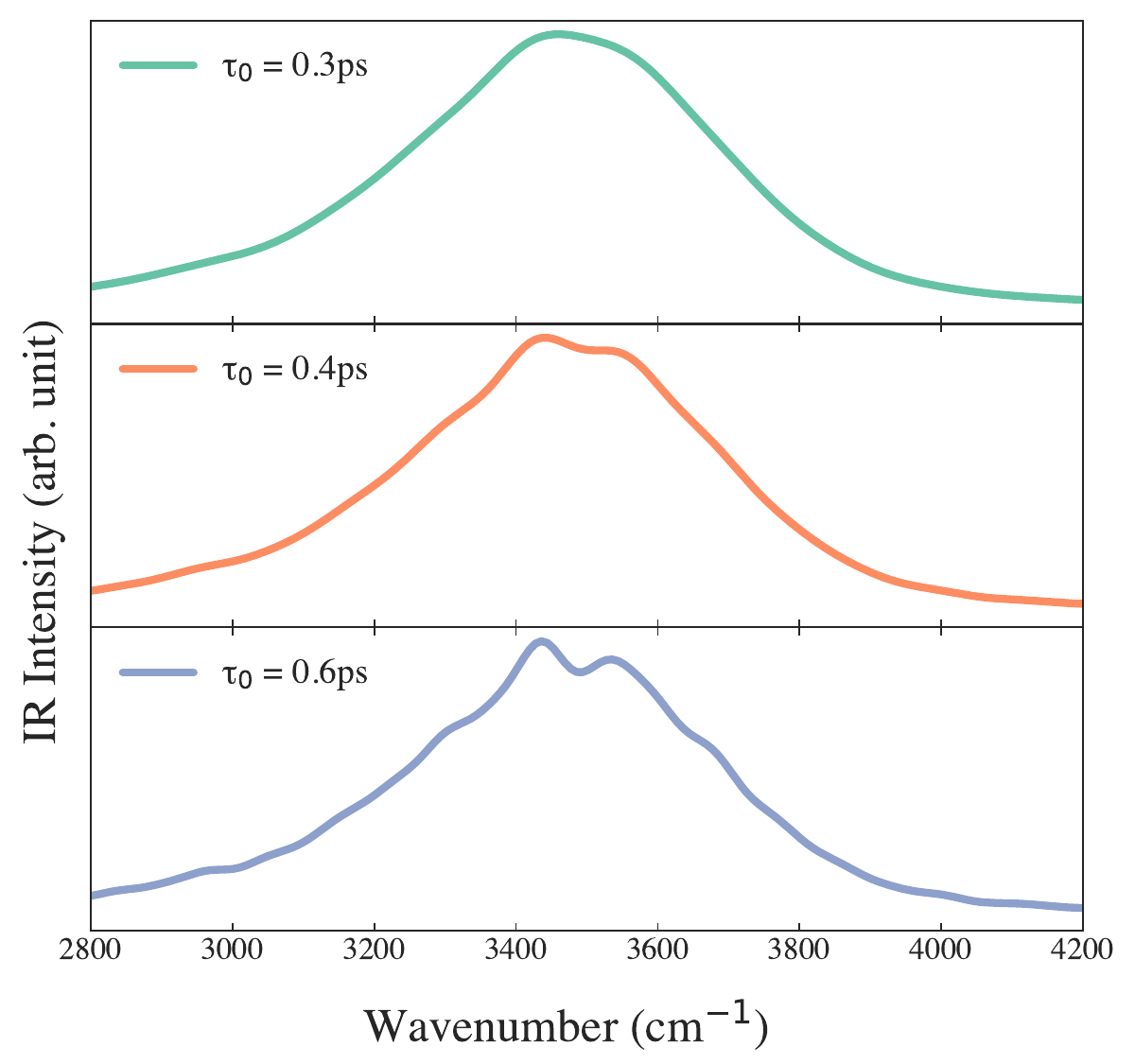}
   \caption{Comparison of the IR spectra in the OH-stretch region for bulk water at 298K with Hann window cutoffs of 0.3 ps, 0.4 ps, and 0.6 ps.}
   \label{fig:tau_compare}
\end{figure*}

\clearpage
\begin{figure*}[!h]
    \addcontentsline{toc}{figure}{Figure \ref{fig:gauss_fit}. Deconvolution of the O–H stretch peaks of the air-water interface from different PI methods.}
   \centering
   \includegraphics[width=0.7\linewidth]{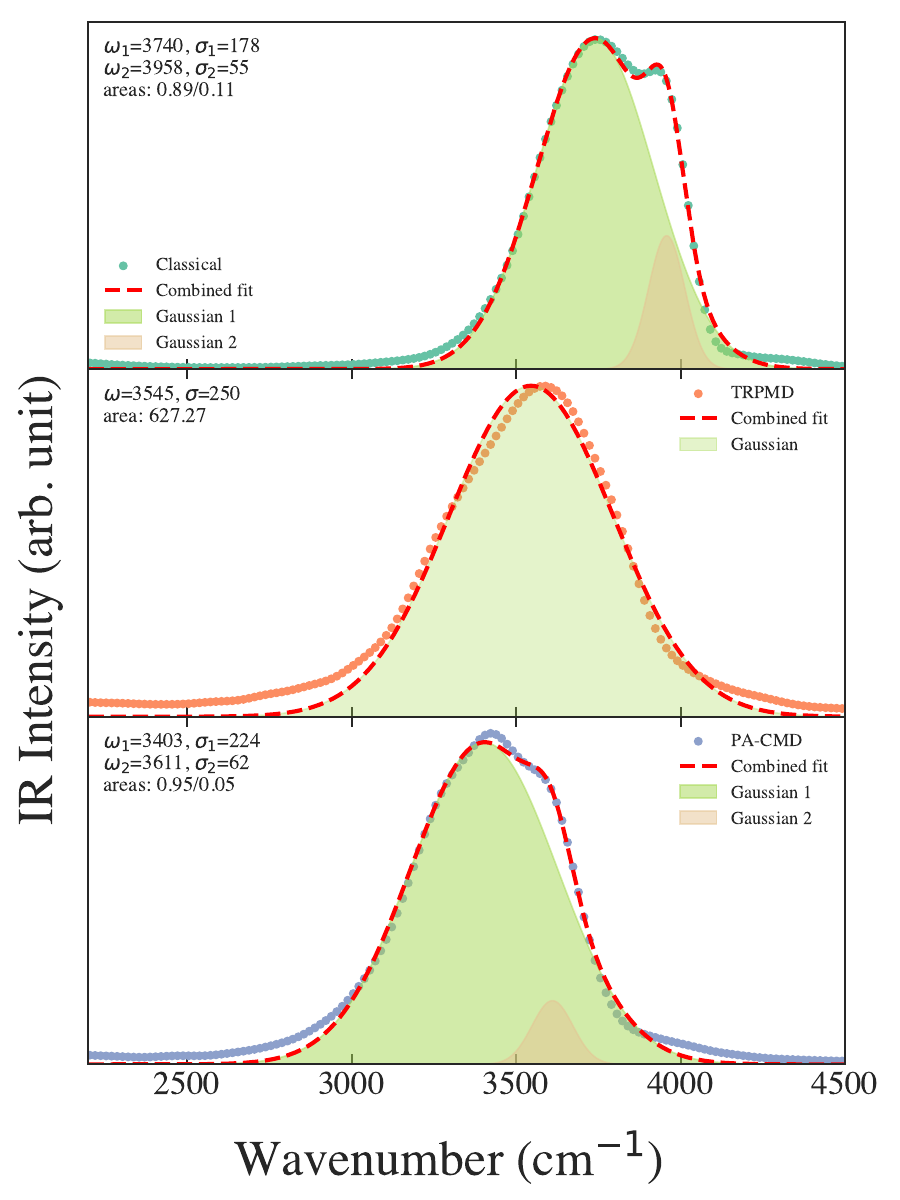}
   \caption{The deconvolution of the O–H stretch peaks of the air-water interface system calculated using different methods.}
   \label{fig:gauss_fit}
\end{figure*}

%\clearpage
%\begin{figure*}[!h]
%    \addcontentsline{toc}{figure}{Figure \ref{fig:simulaiton-box-compare}. Comparison between simulation box size during MD simulation.}
%   \centering
%   \includegraphics[width=0.99\linewidth]{example-image-a}
%   \caption{Comparison between simulation box size during MD simulation.}
%   \label{fig:simulaiton-box-compare}
%\end{figure*}

%\clearpage

%\begin{figure}[!t]
%    \addcontentsline{toc}{figure}{Figure \ref{fig:simulaiton-box-compare}. MD, TRPMD, and PA-CMD simulated IR spectra of bulk water.}
%    \centering
%    \includegraphics[width=0.80\linewidth]{figs/ir-bulk-4500.png}
%    \caption{4G-HDCNNP calculated MD, TRPMD, and PA-CMD simulated IR spectra of bulk water compared to the experimental spectra.\cite{augBertie1996}}
%    \label{sifig:ir-bulk}
%\end{figure}

\clearpage
\bibliography{../bib}